\documentclass[twocolumn,preprintnumbers,amsmath,amssymb,showpacs]{revtex4}
\usepackage{graphicx}% Include figure files
\usepackage{dcolumn}% Align table columns on decimal point
\usepackage{bm}% bold math
\usepackage{times}
\usepackage{multirow}
\usepackage{rotating}

\begin{document}
\title{Quantum switching networks for perfect qubit routing}% Force line breaks with \\
\author{Christopher Facer}
%\affiliation{Physics Department, Macquarie University, Sydney, NSW 2109, Australia}
\author{Jason Twamley}
\author{James Cresser}\affiliation{Centre for Quantum Computer Technology, Physics Department, Macquarie University, Sydney, NSW 2109, Australia}
\date{Sept 30, 2006}
\begin{abstract}
We develop the work of Christandl {\it et al.} [M. Christandl, N. Datta, T. C. Dorlas, A. Ekert, A. Kay, and A. J. Landahl, Phys. Rev. A {\bf 71}, 032312 (2005)],  to show how a $d-$hypercube homogenous network can be dressed by additional links to perfectly route quantum information between any given input and output nodes in a duration which is independent of the routing chosen and, surprisingly, size of the network.

%present a quantum network (the $d-$dimensional hypercube), which achieves perfect qubit transport between antipodes of the hypercube. 

%We show that this basic quantum network fits into a class of quantum networks described by Cayley graphs which allow for pre-determined routing of quantum information: quantum information initially presented at an ``input node'' of the graph is routed through the network to emerge at a unique (and pre-set), ``output node''. These Cayley quantum routing networks are made from the basic Hypercube network with additional, long-range links switched on. We show that the transport fidelity between any ``input node'' and ``output node'' is perfect in all cases and moreover,  the time duration for the routing is independent of the separation between input and output nodes and thus, surprisingly, of the size of the routing network.
\end{abstract}

\maketitle
\indent The study of quantum networks has potential applications in many areas of quantum information science such as quantum communications within multi-party quantum protocols (quantum cryptography, quantum secret sharing etc.), and also within  quantum computer architectures. In this paper we focus on a specific class of quantum network, called the ``dressed hypercube,'' as a means for routing quantum information between nodes within the network. These networks are similar to other networks capable of transporting qubits perfectly but they have the added quality that the destination of the qubits may be controlled by an external user.  

Any implementation of quantum information processing which is not based on optical qubits will require a mechanism for transporting qubits between gates and processors. There have been several theoretical proposals for qubit transport which are based on a chain of spin-half particles which are coupled by Heisenberg or XY interactions.  The first proposal \cite{Bose2003}, was a homogenous chain of particles coupled by homogeneous, nearest-neighbour interactions. A qubit is encoded at one end of the chain and the system evolves. The probability of retrieving an encoded qubit from the destination end of the chain was found to diminish as the length of the chain increased. Later, \cite{Christandl2004}, found that chains of any length were able to transport qubits perfectly but only if the coupling between neighbouring particles was inhomogeneous and carefully engineered in such a way as to be strong at the middle of the chain and weaker towards the ends of the chain. More recently it was shown that by relaxing the degree of control to encompass global addressing of all the particles in the chain also allows for perfect transport, \cite{Raussendorf2005,Fitzsimons2006}.

Some work has also been done in examining the properties of quantum networks which are more complicated than linear chains. It was found in \cite{Christandl2005}, that hypercubic networks of any dimension are capable of transporting qubits between pairs of anitpodal nodes. In fact, if only a single pair of input and output nodes are considered, the hypercube reduces to the inhomogeneous chain. Below we show that by introducing additional links into a hypercubic network in a specific way, the destination node of a qubit can be changed. Thus, if a user were able to choose which extra links in the network were ``switched on'' they will be able to route a qubit to any desired destination within the network and in a duration which is independent of the network size.

We take a quantum network to be a collection of $N$ spin-half particles, each of which is situated on one of the nodes of an undirected graph $G:=\left\{V(G),E(G)\right\}$, made up of nodes $V(G)$, and connecting edges $E(G)$. The edges of the graph represent the allowed couplings between these particles, i.e. if two nodes $i$ and $j$ are connected on the graph, then $(i,j)\in E(G)$, and the two particles are coupled by an XY interaction $H_{ij}=J_{ij}[\hat{\sigma}_i^x\hat{\sigma}_j^x+\hat{\sigma}_i^y\hat{\sigma}_j^y]$. In what follows we will chose the coupling strengths $J_{ij}=1$. The total Hilbert space of the system is ${\cal H}_G=\otimes_{k\in V(G)}\,{\cal H}_k=(\mathbb{C}^2)^N$, where $N=\#V(G)$, the number of nodes in $G$.

The \emph{adjacency matrix}, $A(G)$, of a graph $G$, captures all of the connections of the graph and is defined by:
\begin{equation}
A_{ij}(G) = \left\{\begin{array}{cc}
							1 & \textrm{ if $i$ is connected to $j$}\\
							0 & \textrm{if $i$ and $j$ are not connected}.
							\end{array}\right.
\end{equation}
\label{adjacencymatrix}

Using the adjacency matrix we can write down the Hamiltonian for the network of interacting particles as:

\begin{equation}
\hat{H}_{XY} = \frac{1}{2}\sum_{i,j}A_{ij}(\hat{\sigma}_{x}^{i}\hat{\sigma}^{j}_{x} + \hat{\sigma}_{y}^{i}\hat{\sigma}^{j}_{y}), \label{Ham1}
\end{equation}

\noindent where the factor of $\frac{1}{2}$ accounts for the fact that the summation includes all pairs of interacting particles twice.
Crucially, Hamiltonians of the form (\ref{Ham1}), (and more generally with any additional terms of the form $\hat{\sigma}_z^i\hat{\sigma}_z^j$, which then encompass Heisenberg coupled Hamiltonians), conserve the total $z$-spin of the particles in the network, i.e. $[\hat{H}, \hat{\sigma}_{z}^{tot}] = 0$, where $\hat{\sigma}_{z}^{tot} = \sum_{n=1}^{N}\hat{\sigma}_{z}^{n}$. Thus the evolution occurs in separate invariant eigenspaces of the total Hilbert space, each labeled by the eigenvalue of $\hat{\sigma}_z^{tot}$.  In the case where we allow a single excitation the evolution can be easily studied in a basis of $N$ node states. We will represent these single excitation basis states for the single excitation subspace as $|k\rangle$, where all spins are down except the $k^{th}$, spin which is up. Further, in this restricted single excitation case the XY Hamiltonian (\ref{Ham1}), is proportional to the adjacency matrix of the network.  Similarly, for an Heisenberg coupled Hamiltonian, the total Hamiltonian becomes proportional to a related matrix known as the \emph{Laplacian} of the network.

This model now allows us to explore the quantum dynamics of a particular network via the adjacency matrix of the network. In the case of networks based on the hypercube and the ``dressed hypercube'', (which we define below), we will show that the time evolution of the system performs a permutation of the states of the nodes in the  network, and it does so periodically. This means that any qubit encoded on an ``input'' node becomes swapped with the spin state of the particle at an ``output'' node, effectively transporting the qubit through the network. Moreover, the permutation that is performed can be changed by dressing the underlying hypercube network in different ways,  and this allows for the routing of the input qubit from any given input node to any given output node perfectly. 

The paper is organised as follows. In Section \ref{A} we review the work of \cite{Christandl2005}, to show how single-link and double-linked Hypercubes can admit perfect quantum transport. We also review their construction of a more general network which admits perfect transport between very particular antipodal nodes. In section \ref{B} we expand their analysis to Cayley graphs, or ``dressed hypercubes'', which are basically hypercube networks with specifically chosen additional links. We prove that the structure of the adjacency matrices of these dressed networks can always be written as a Kronecker product between two simple matrices. In section \ref{C} we develop methods  to characterise the spectrum of these dressed networks and thus determine the quantum dynamics on these networks. We show that the evolution, at specific times, permutes the  quantum states of the nodes in the network in the single excitation subspace. We further find that the times at which the evolution corresponds to a permutation are independent of the specific permutation and $N$, the number of nodes in the network.  We finally show that these new class of perfect transport networks do not fit into the very general category discovered in \cite{Christandl2005}.

\vspace{-.5cm}\section{State transfer in Hypercubes}\label{A}
\vspace{-0.2cm}As mentioned above, our router consists of dressing a basic hypercube network with extra links. Before examining these dressed hypercubes it is instructive to review the perfect quantum transport of single excitations between antipodes on a $d$-dimensional hypercube. We follow \cite{Christandl2005}, but later on we develop another proof which we can also apply to dressed hypercubes. One considers the network initialised with one overall excitation localised at one node, $|A\rangle$, which evolves over the network and recoheres at another node $e^{\i\phi}|B\rangle=\exp(-i H_G \tau)|a\rangle$, up to a global phase $\phi$. One now assumes that the network $G$ appears identical from both the viewpoints of nodes $A$ and $B$, i.e. we say that the network is {\em mirror symmetric}, when viewed by $A$ and $B$. Under this special condition the subsequent evolution for a time $\tau$ will cause the wavefunction to recohere back again at $A$ (up to a global phase), and thus $|\langle A|\exp(-2iH_G\tau)|A\rangle|=1$. One can show that for this to be possible then $H_G$ must possess an energy eigenvalue spectrum $E_k$, such that the difference ratios are all rational fractions, i.e. $(E_i-E_j)/(E_{i^\prime}-E_{j^\prime})\in\mathbb{Q},\;\forall\,(i,j,i^\prime, j^\prime, i^\prime\ne j^\prime)$. In \cite{Christandl2005} they consider the energy spectra of homogeneously coupled nearest-neighbor spin chains of length $N$ and prove that the above eigenvalue condition for perfect transfer is only possible for single and double linked chains, i.e. when $N=2, 3$. To get perfect transport over larger graphs they consider the Cartesian product of small (single link), perfect transfer chains. Considering the Cartesian product $L=G\times H$, and denoting the eigenvalues of the component graphs, $\{\lambda_i(G), 1\le i\le|V(G)|\}$, and $\{\lambda_j(H), 1\le j \le |V(H)|\}$, then $\lambda_k(L)=\lambda_i(G)+\lambda_j(H)$. To show this one considers how the adjacency representation of the Cartesian product is formed $A(G\times  H)=A(G)\otimes \mathbb{I}_{|V(H)|}+\mathbb{I}_{|V(G)|}\otimes A(H)$. Using these properties one can show that the eigenspectrum condition for perfect transport is satisfied for a graph $G^d=G\times G\cdots\times G$, if $G$ itself does. The authors in \cite{Christandl2005}, also presented another method of constructing a perfect transport network via its reduction to a 2D column representation. More specifically they consider graphs that can be arranged into columns of nodes and where edges connect only adjacent columns. Each node in a column $i$ possesses an identical number of ``backward links'' to nodes in the previous column $i-1$, and an identical number of ``forward links'', to nodes in the next column $i+1$. Further there are no links connecting nodes within a column $i$. With this construction they find examples of perfect transport (via a correspondence with hypercubic graphs), and then quantum transport on a one-dimensional spin chain with ``engineered'' coupling strengths. Below we will show a new construction of a perfect transport spin network which is not of this (already quite general), columnar form.

\vspace{-0.5cm}\section{Cayley Networks}\label{B}
\begin{figure}
 \begin{center}
\setlength{\unitlength}{1cm}
\begin{picture}(5,4.5) 
\put(7.4,-1){\begin{rotate}{90}\includegraphics[height=10cm,width=6.5cm]{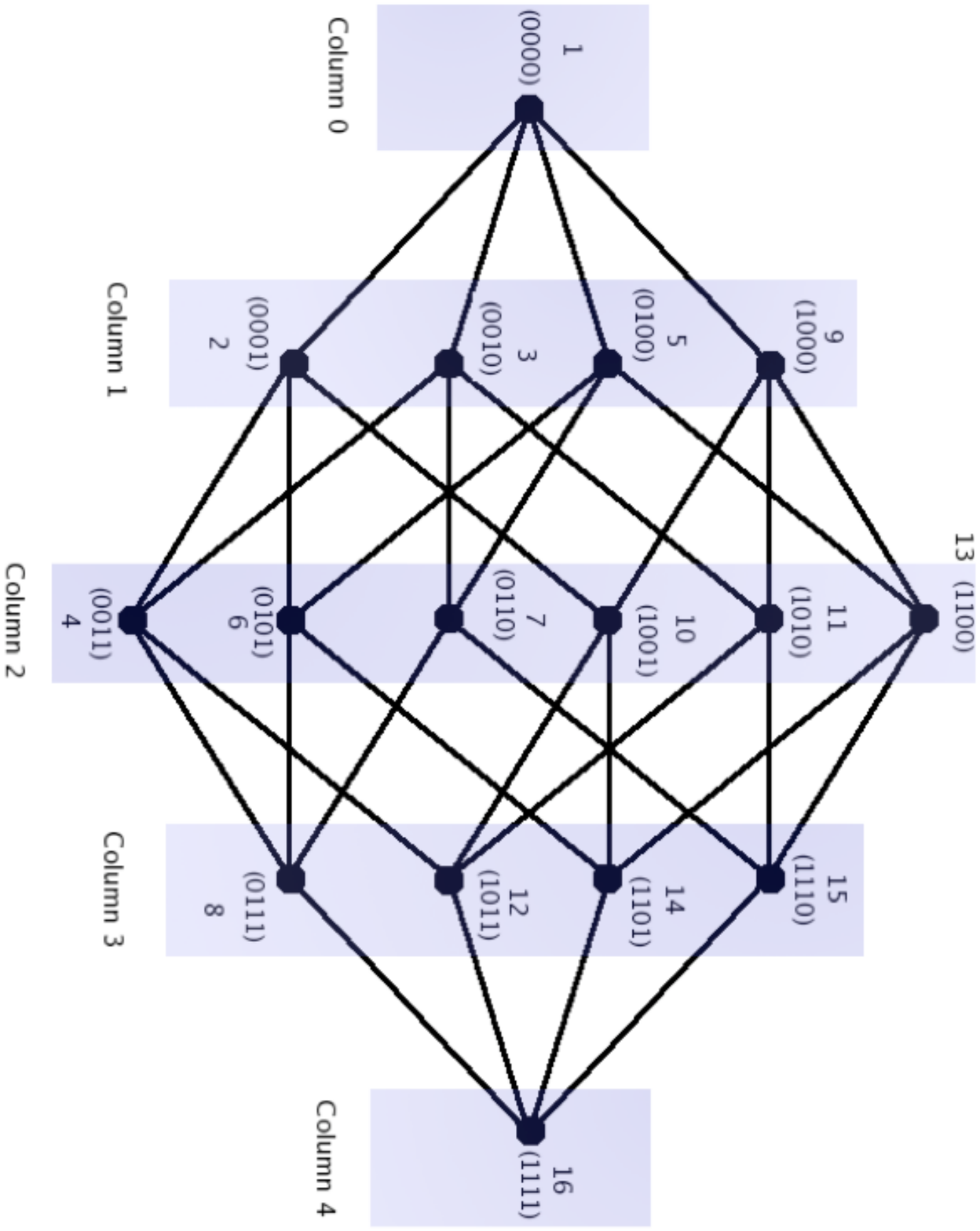}\end{rotate}}
\end{picture}\end{center}
\caption{Illustration of the binary labeling of the nodes of a $d$-dimensional hypercube, (here $d=4$). Nodes with equal Hamming weight can be arranged into columns. Edges only have unit Hamming length.\label{fig:fig1}}
\end{figure}

\vspace{-0.2cm}We now introduce another method of constructing hypercube networks which also encompasses more general ``dressed'' hypercube networks called {\em Cayley networks}. By going to a binary labeling of the nodes we can find a  group representation decomposition of the adjacency matrix of Cayley networks which allows us to prove the perfect transfer properties of hypercubes and dressed hypercubes. We first define the Caley network ${\cal C}ay(G,S)$, of a finite group $G$ (with identity element $e$), with $S\subset G$, being a generating set of $G$, to be the network ${\cal C}ay(G,S)=\{ V, E\}$, where $V$, the vertex set, corresponds to the elements of $G$, while the edge set $E=\{(x,y)\,|\,y=xg$, for some $g\in S \}$. We now consider $G=\mathbb{Z}_2^d$, an elementary commutative group under addition modulo 2, of order $2^d$ (here $\mathbb{Z}_2=(0,1)$). The identity element is $e\equiv \{(0,0,\cdots 0)\}$, and we consider the family of generating subgroups $S^l_d=H^1_d\cup H^2_d$, where $H^1_d=\{(x_1,x_2,\cdots,x_d)\in \mathbb{Z}_2^d|$ only one of $x_1,\cdots,x_d$ is 1$\}$, and $H^2_d=\{(x_1,x_2,\cdots,x_l,0^{d-l})\,|\,(x_1,x_2,\cdots,x_l)\in \mathbb{Z}_2^l\backslash e\}$. One can compute that $|H^1_d|=d$, while $|H^2_d|=2^l-1$, and that $|S^l_d|=2^l+d-l-1$.  We can form the Caley binary network $\mathbb{Z}_2^d(l)\equiv {\cal C}ay(\mathbb{Z}_2^d,S^l_d)$. Initially we will set $l=1$, to consider $d-$dimensional hypercubes and, as an illustration we set $d=3$, whereupon $H^2_3=\{(1,0,0)\}$, $H^1_3=\{(1,0,0),(0,1,0),(0,0,1)\}$, and $|S^1_3|=3$. The $2^3$ nodes of $\mathbb{Z}_2^3(1)$, are connected via the Cayley edge relation $g_j=s_k\oplus g_i$, where $g_l\in V(G)$, and $s_k\in S^1_3$, and the group multiplication operation is addition modulo two. Thus the node $(0,0,0)$ is connected to the nodes $(0,1,0),\, (0,1,0),\, (0,0,1)$, while the node $(1,1,0)$ is connected to the nodes $(0,1,0),\, (1,0,0),\,(1,1,1)$ (see Fig. \ref{fig:fig1}). From this construction it is clear that that $\mathbb{Z}_2^d(l)$, are regular networks, i.e. the number of edges meeting at a node is identical throughout the network. The regular hypercubic networks, $\mathbb{Z}_2^d(l=1)$, exhibit $d$ edges per node and it is clear from simple examples that one can arrange the nodes into columns  labeled by the Hamming weights of their $\mathbb{Z}_2^d(l=1)$, representation, and connected by edges of unit length Hamming distance. Such a column arrangement satisfies the general construction of \cite{Christandl2005}, and thus, by their proof, exhibits perfect quantum transport between nodes $(0,0,\cdots, 0)$, and $(1,1,\cdots,1)$.  We now exhibit an alternative proof which will be used later when $l\ne 1$.

We make use of the following decomposition of the overall adjacency matrix of the graph $G$,  into components within the generating set: 
\begin{equation}
A(G)=\sum_{a\in S^l_d} \,\rho(a)\;\;,\label{decomp}
\end{equation}
  where $\rho$ is the fundamental adjacency representation of an element $a$, in the generating set $S^l_d$, given by
%\begin{equation}
$\rho( a= (x_1,x_2,\cdots,x_d))=X_1\otimes X_2\otimes \cdots \otimes X_d$,
%\end{equation}
where 
\begin{equation}
X_i=\left\{\begin{array}{ll} \mathbb{I}_2 & {\rm if}\; x_i=0\;\;,\\
C&{\rm if}\; x_i=1\;\;,\end{array}\right. 
\end{equation}
and where $\mathbb{I}_2=(\begin{array}{cc} 1 & 0\\0 & 1\end{array})$, and $C=(\begin{array}{cc} 0 & 1\\1 & 0\end{array})$, is the fundamental matrix representation of a swap permutation. Thus, for a $d=3$ (hyper)cube, $A(G)=C\otimes \mathbb{I}\otimes \mathbb{I}+ \mathbb{I}\otimes C \otimes \mathbb{I}+ \mathbb{I}\otimes \mathbb{I}\otimes C$.  We further note that all matrices of the form $\{ X_1 \otimes X_2\otimes \cdots \otimes X_d$: each $X_j$ is either $\mathbb{I}_2$ or $C$$\}$, form a much larger group under matrix multiplication which we will call the {\em Kronecker Product Group} of dimension $d$. The eigen-structure of $A(G)$, can now be broken down via the decomposition (\ref{decomp}), as the eigenvectors of $D=A\otimes B$, are of the form $|a\rangle\otimes |b\rangle$, where $|a\rangle$, is an eigenvector of $A$, and similarly for $|b\rangle$. Thus the eigenvectors of the adjacency matrix of the $d-$hypercube must then take the form $|\lambda\rangle=|x_1\rangle\otimes |x_2\rangle\otimes\cdots\otimes |x_d\rangle$, where $|x_i\rangle =|+1\rangle$, or $|-1\rangle$. Given that $A(G)=C\otimes \mathbb{I}\otimes \mathbb{I}+ \mathbb{I}\otimes C \otimes \mathbb{I}+\cdots+ \mathbb{I}\otimes \mathbb{I}\otimes C$, with $d$ terms for the $d-$hypercube, we can see that the maximal eigenvector is $|\lambda_d\rangle=|1\rangle\otimes |1\rangle\cdots|1\rangle$, with eigenvalue $d$. Since switching an $x_i$, in $|\lambda\rangle$, from $x_i:1\rightarrow -1$, reduces the overall eigenvalue by $2$, we can classify the ladder of eigenvalues of $A(G)$, into three categories, (H1) Maximal eigenvector, with an eigenvalue $\lambda_{H1}^d=d$, (H2) $n^{th}$ even group of eigenvectors with $2n$, $|-1\rangle$, components in each eigenvector, with an overall eigenvalue of $\lambda_{H2}^d(n)=d-4n$, (H3) $n^{th}$ odd group of eigenvectors with $(2n-1)$, $|-1\rangle$, components in each eigenvector with an overall eigenvalue of $\lambda_{H3}^d(n)=d-4n+2$. The eigenvalues form a ladder from $+d, +d-2, \cdots, -d+2, -d$, with the minimum eigenstate $|\lambda_{-d}\rangle=|-1\rangle\otimes\cdots\otimes|-1\rangle$. Crucially, we now note that given $d$, we can always choose an integer $k$,  such that $d-k$, $\lambda_{H2}^d(n)-k$, and $\lambda_{H3}^d(n)-k+2$, are all multiples of 4. This fact now allows us to express the quantum evolution of a state initially localised on node $|m\rangle=\sum_{j=1}^Nc_j|\lambda_j\rangle$, when decomposed over the eigenstates of $A(G)$, over a period of time $\tau=\pi/2$, to be
\begin{eqnarray}
\hat{P}|m\rangle&=&e^{-i\hat{H}\pi/2}|m\rangle=\sum_{j=1}^N\, c_je^{-i\lambda_j\pi/2}|\lambda_j\rangle\;\;,\\
&=&e^{-ik\pi/2}\sum_{j=1}^N\, e^{-i(\lambda_j-k)\pi/2}|\lambda_j\rangle\;\;.
\end{eqnarray}
From the above we can see that the phase factor within the sum will take the value $+1\{-1\}$ for all eigenvectors in the categories (H2)$\{$(H3)$\}$, and thus $\hat{P}^2=\mathbb{I}$, and that all of the eigenstates are either symmetric or anti-symmetric under $\hat{P}$.  From our category analysis above there are an equal numbers of symmetric and antisymmetric eigenstates. To show that $\hat{P}$ generates a permutation of the nodes, and in particular, $\hat{P}=C^{\,\otimes\,d}$, which swaps the quantum state between antipodes, we recall that for $d-$hypercubes the Hamiltonian/adjacency matrix is a sum of matrices from the Kronecker product group of dimension $d$. Since, $\hat{P}=\exp(ik\pi/2)\exp(-i\hat{H}\pi/2)=\sum_{j=0}^\infty\, d_j (\hat{H})^j$, this power series expansion of the operator exponential must also be expressible as a sum of elements of the $d-$Kronecker product group. However, since $\hat{P}^2=\mathbb{I}$, then the sum must only contain one term. The only such term which possesses the appropriate symmetry conditions (an equal number of symmetric and antisymmetric $|\lambda_j\rangle$), is $\hat{P}=C^{\,\otimes d}$.  Thus by direct computation we have shown that the Hamiltonian, evolved for a duration $\tau=\pi/2$, yields a permutation of the single excitation subspace exchanging antipodes of the hypercube.
\vspace{-0.5cm}\section{Dressed Hypercubes}\label{C}
\begin{figure}
 \begin{center}
\setlength{\unitlength}{1cm}
\begin{picture}(5,2.5) 
\put(-2.5,-2.7){\includegraphics[height=8cm,width=6cm]{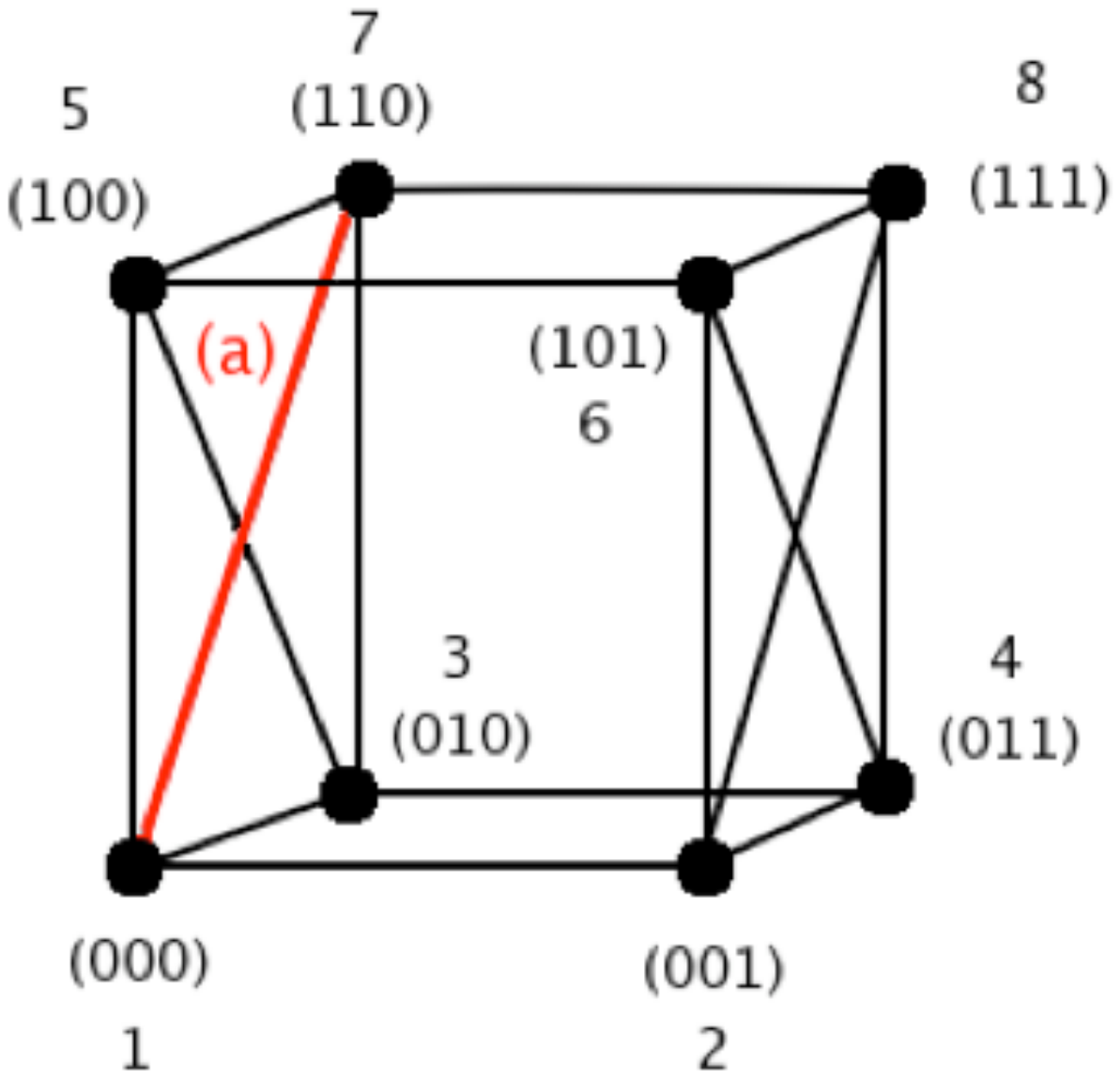}}
\put(1.7,-2.7){\includegraphics[height=8cm,width=6cm]{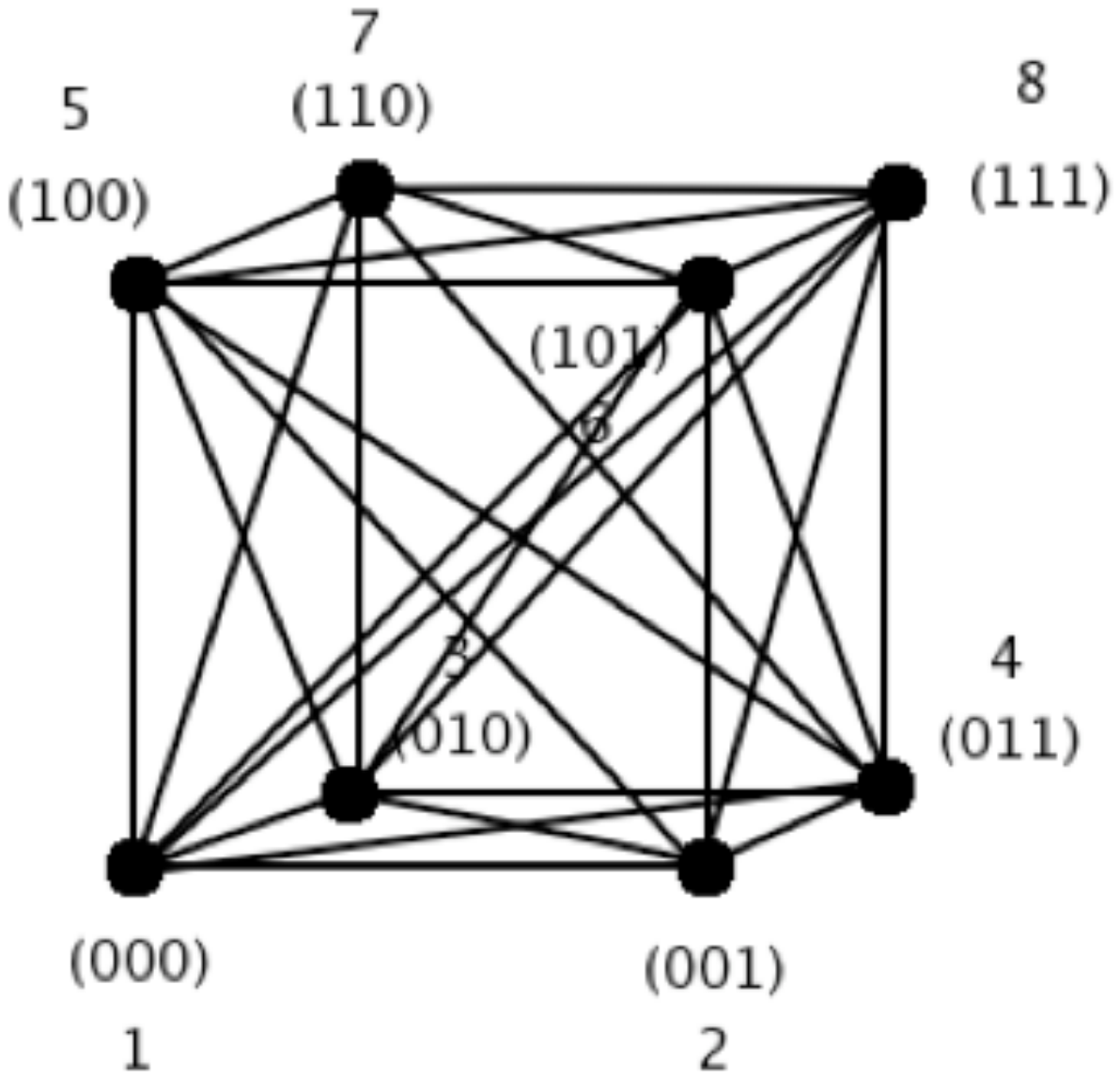}}
\put(-1.6,2.5){(i)}
\put(2.6,2.5){(ii)}
\end{picture}\end{center}
\caption{(i) The graph of the dressed hypercube $\mathbb{Z}_2^3(2)$. The link labeled (a) is the link generated 
from $(000)$ by the  generating set element $(110)$. (ii) The graph of the dressed hypercube $\mathbb{Z}_2^3(3)$, a complete graph with 8 nodes.\label{fig:fig2}}
\end{figure}

\vspace{-0.2cm}We now consider the Calyey networks where $l>1$.  From the definition of $\mathbb{Z}_2^d(l)$, the set of generators now expands introducing new edges into the network, e.g. in $d=3, l=2$, we obtain the single extra generator $(1,1,0)$ (see Fig. \ref{fig:fig2}). However the Kronecker decomposition of $A(G)$, into 
products of $\mathbb{I}_2$ and $C$ still holds and thus the eigenstates of $A(G)$ are composed up of Kronecker products of $|\pm 1\rangle$. We again use this to map out the eigenspace of the overall adjacency matrix. As before we find a ladder of eigenvalues ranging from the maximum 
$\lambda_{max}=2^l+d-l-1$, (also the degree of the network), to a minimum, but now crucially $
\lambda_{min}\ne  -\lambda_{max}$. One can follow the same arguments as for the hypercube to find that eigenvectors fall into two categories, even and odd, which determine their symmetries under the quantum evolution operator $\hat{P}=\exp(-i\hat{H}\pi/2)$. We also find $\hat{P}^2=\mathbb{I}$, and thus $\hat{P}$, is again expressible as one element of the Kronecker product group but this time there are unequal numbers of symmetric and antisymmetric eigenstates of $\hat{H}\sim A(G)$, and we are led to identify $\hat{P}$, with a particular permutation operator of the nodes of the network.  The choice of permutation is thus dictated by the numbers of even and odd  eigenvectors of $A(G)$. More specifically, we find that the eigenvalues of $A(G)$, ensure that the permutation operation be such that eigenvectors with even numbers of $|-1\rangle$'s in the last $d-l$ terms of their Kronecker product expansion are symmetric whilst those with an odd number are antisymmetric, e.g. for $d=3,\, l=2$, the symmetric eigenstates are those with a $|1\rangle$ in the rightmost slot.  When $l=1$,  categorising these eignvectors  was relatively straightforward but this becomes more complex when $l>1$. It turns out useful to break up the terms in the Kronecker product sum expansion of $A(G)$ into class $S_1$: sums of elements generated by generators with $d-l$, $0$'s at the end, which corresponds to Kronecker terms which each end with $\mathbb{I}^{\otimes\,d-l}$. The remaining class $S_2:$ corresponds to all other elements of the Kronecker product group not in $S_1$. For examples, $d=3,\, l=2$, $S_1$ contains $A_1=C\otimes \mathbb{I}_2\otimes \mathbb{I}_2+ \mathbb{I}_2\otimes C \otimes \mathbb{I}_2+C\otimes C\otimes  \mathbb{I}_2$, while $S_2$ contains, $A_2= \mathbb{I}_2\otimes \mathbb{I}_2\otimes C$. One can compute the eigenvalue for any given eigenvector by considering the terms in the adjacency expansion in the two categories $S_{1,2}$.  Each term in the adjacency sum expansion will contribute $\pm 1$, to the overall eigenvalue.
For each 
term in this expansion, if the number of $C$'s (one for each $|-1\rangle$, in the eigenvector), in the product expansion of that term is even(odd) then this term contributes $+1(-1)$ to the overall eigenvalue.
\begin{figure}
 \begin{center}
\setlength{\unitlength}{1cm}
\begin{picture}(5,3.5) 
\put(-2.2,-1.2){\includegraphics[height=5cm,width=9.5cm]{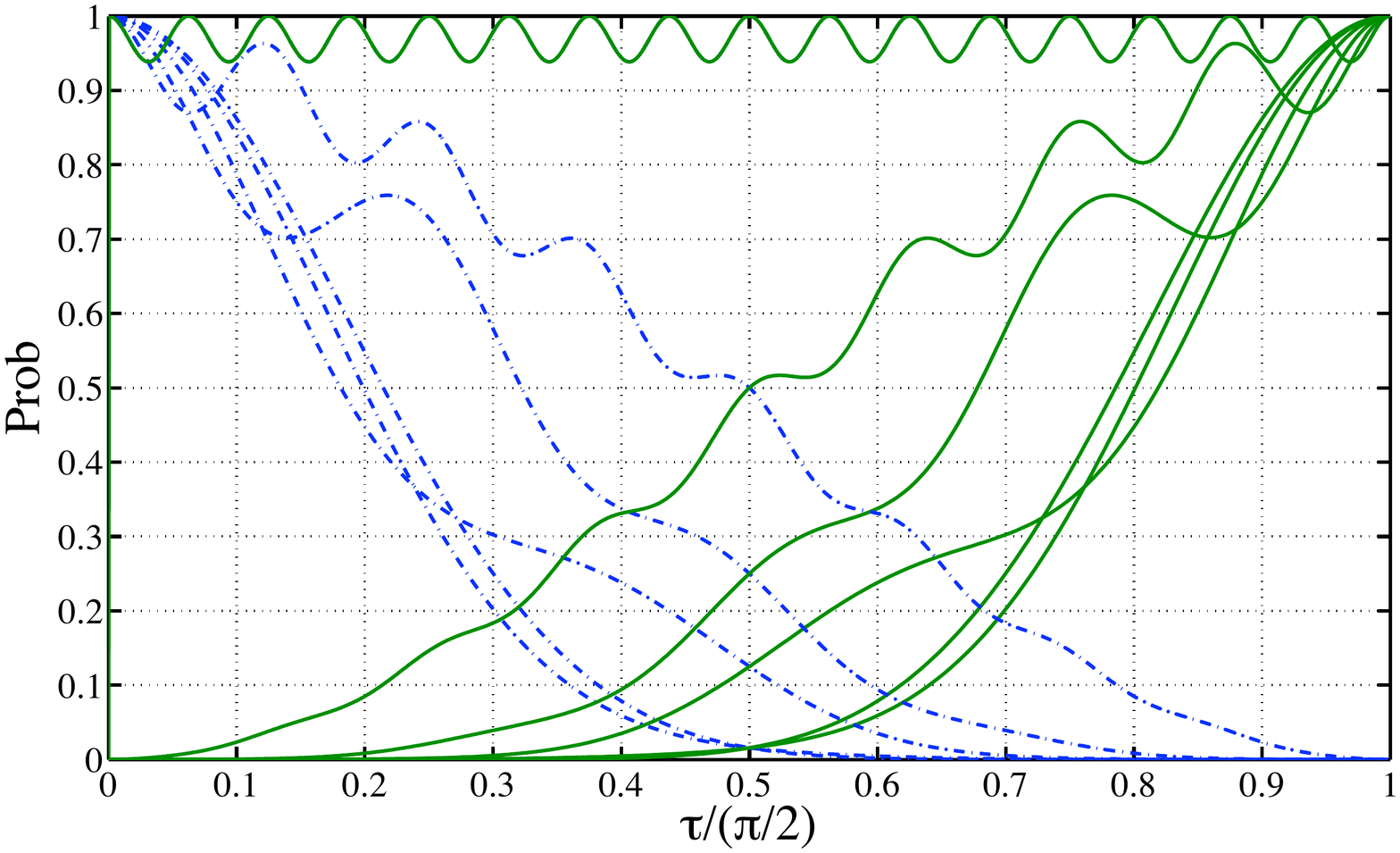}}
\put(4.6,1.3){\small $1$}
\put(4.4,.5){\small $2$}
\put(3.4,.5){\small $3$}
\put(3.8,1.1){\small $4$}
\put(3.8,1.88){\small $5$}
\put(3.2,2.6){\small $6$}
\end{picture}\end{center}
\caption{Graphs of $|\langle 0|U(\tau)|0\rangle|^2$, (dotted), and $|\langle T|U(\tau)|T\rangle|^2$, (solid), for $\mathbb{Z}_2^6(l)$, where $l=1,...,6$, and $T$ is the node targeted for perfect transport. 
%At $\tau/(\pi/2)=1$, the target node receives the qubit perfectly. 
\label{fig:fig3}}
\end{figure}

Using this, and after some work, one can show that the contribution of all terms in the sum expansion of the eigenvector in the category $S_1$ contribute a factor of $2^l$, to the overall eigenvalue if that eigenvector contains no $|-1\rangle$, in the first $l$ slots, and contributes only $-1$ otherwise. Similarly the terms in the expansion in $S_2$ give an overal contribution to the eigenvalue of $-P_1+P_2$, where $P_1=\#(|-1\rangle)$, $P_2=\#(|+1\rangle)$, (the numbers of these), in the rightmost $d-l$ slots of the eigenvector. Using these one can see that the eigenvalues for $\mathbb{Z}^d_2(l)$, form a ladder and split into {\em even} $\lambda_{even}^d(n)=2^l+d-l-1-4(n-1)$, and {\em odd} $\lambda_{odd}^d(n)=2^l+d-1-4(n-1)-3$, subsets. We can again find an integer $k$ such that $\lambda_{even}^d(n)-k$, and $\lambda_{odd}^d(n)-k+2$, are all multiples of 4 and thus all these eigenstates have eigenvalues $\pm 1$, under $\hat{P}$. However we find now that if the eigenvector, $|\lambda_j\rangle$, has an even number of $|-1\rangle$'s in the last $d-l$ slots, then it is symmetric under the permutation, $\hat{P}|\lambda_j\rangle=|\lambda_j\rangle$, otherwise it is antisymmetric, i.e. $\hat{P}|\lambda_j\rangle=-|\lambda_j\rangle$. As $\hat{P}^2=\mathbb{I}$, the only single term in the Kronecker Product group possible which respects these symmetries is $\hat{P}=\mathbb{I}^{\otimes\, l}\otimes C^{\otimes\, d-l},\;1<l\le d$, e.g. for $d=3,\,l=2$, $\hat{P}=\mathbb{I}\otimes \mathbb{I}\otimes C$. It is interesting to note that this predicts that for $l=d$, (a fully connected graph), $\hat{P}=\mathbb{I}^{\otimes\, d}$, the identity and the excitation, after evolving for a period $\tau=\pi/2$, returns perfectly to its starting node. When $d=3$, and $l=1$, we get the permutations $(1,8)(2,7)(3,6)(4,5)$, and when $l=2$ we obtain $(1,2)(3,4)(5,6)(7,8)$, (see Fig. \ref{fig:fig3} for $\mathbb{Z}_2^6(l)$). By combining these permutations and those obtained by considering the perfect transport on rotated Calyey networks (where we consider dressing the basic hypercube following any rotation of the hypercube about some axis of symmetry), we can route a single excitation from any given input node of the network perfectly to any given output node. The time for such routing can be computed precisely given the input/output nodes. Further, the Cayley networks $\mathbb{Z}_2^d(l)$, for $l>1$, no longer satisfy the general columnar construction of \cite{Christandl2005}, as the new generators add links connecting nodes separated by Hamming distances greater than unity.  We expect that this protocol for perfect quantum routing by dressing the basic hypercube network with additional links may be of use in quantum computation, cryptography and communication.

This work has been supported by the EC IST QAP Project Contract Number 015848.

\bibliography{router1}

\end{document}